\documentclass{elsart}
\usepackage{natbib}

\begin{document} 

\begin{frontmatter}
\title{The Triplet Genetic Code had a Doublet Predecessor}

\author{Apoorva Patel}
\address{
Centre for High Energy Physics and\\
Supercomputer Education and Research Centre\\
Indian Institute of Science, Bangalore 560012, India\\
}
\ead{adpatel@cts.iisc.ernet.in}

\begin{abstract}
Information theoretic analysis of genetic languages indicates that the
naturally occurring 20 amino acids and the triplet genetic code arose
by duplication of 10 amino acids of class-II and a doublet genetic
code having codons NNY and anticodons $\overleftarrow{\rm GNN}$.
Evidence for this scenario is presented based on the properties of
aminoacyl-tRNA synthetases, amino acids and nucleotide bases.
\end{abstract}

\begin{keyword}
Aminoacyl-tRNA synthetases \sep amino acid R-groups \sep genetic code
\PACS 87.14.-g, 87.23.Kg
\end{keyword}
\end{frontmatter}

There exists a broad consensus in biology that evolution, acting through
natural selection on variations produced by genetic mutations, has brought
living organisms to their present state, and would continue to take it still
further. Evolution attempts to explain the highly complex mechanisms of life,
observed in present day organisms, as arising from accumulation of small
changes on simpler predecessors and over a long time scale. Among the many
possible changes in a working system, most are harmful and a beneficial
change occurs only rarely. But natural selection wipes out the undesirable
changes, and amplifies the rare beneficial mutation. This view of evolution
is very well supported by systematic analysis of fossil records and genome
sequences. In this view, it is quite logical to believe that evolutionary
changes can only be incremental, because a large change in a vital part of
life would be highly deleterious \citep{frozen}.
Nonetheless, large rapid changes have occurred during evolution, and two
underlying routes for them have been discovered. One route is duplication
of genes, which allows one copy to carry on the required function while the
other is free to mutate and give rise to a new function. This process has
produced many homologous families of proteins. Another route for increasing
the capability of a particular organism is the import of fully functional
genes developed by a different organism. Indeed, symbiotic transfers of
whole genomes have given rise to organelles, such as mitochondria and
chloroplasts, in eukaryotic cells.

All these advances in understanding evolution have enabled us to construct
a "tree of life", at the root of which lies a prokaryotic proto-cell that
would be the common ancestor of all the living organisms. Despite these
advances, the origin of life itself, i.e. how the proto-cell came about,
has remained a subject shrouded in mystery. We are unable to reconstruct,
with any reasonable measure of confidence, the circumstances prevalent on
earth when the proto-cell came in to being. The fossil records become scanty
as we extrapolate back in time and as the size of organisms decreases.
And evolution itself, due to its optimizing nature, has wiped out traces of
earlier simpler forms of life. The simplest proto-cells that we can track
back life to would necessarily possess sufficient machinery to support
life's fundamental processes---reproduction and metabolism. That would still
require hundreds of genes and thousands of different types of molecules, and
it is too complex a system to have been readily produced from a primordial
soup of organic molecules. How can we bridge the gap? The clues are meagre
and conjectures abound. Our only hope for a solution is to connect the
fragments of information to the physical properties of the ingredients.

Here I focus on one particular aspect of this puzzle, namely how the
languages of genes and proteins arrived at their present structure.
These two languages consist of 4 nucleotide bases and 20 amino acids
respectively, and are connected by a non-overlapping triplet genetic code.
They are universal, and so are expected to be present in the proto-cell,
but they are also too complicated to get established in one go. Many other
nucleotide bases and amino acids exist within cells (e.g. in tRNA, rRNA
and proteins), but they do not participate in these languages, and are
generally synthesized by modifications after transcription/translation.
This fact implies that some optimization criteria have narrowed down the
choice of building blocks for these languages from many possibilities.
Such criteria would involve availability and functional efficiency of
each and every building block of the languages.

Discovery of simpler predecessors to these languages, containing a smaller
number of building blocks, would definitely be a step towards understanding
the mysterious origin of life and development of its complexity. Towards this
goal, many attempts have been made since the discovery of the genetic code
\citep[see for instance:][]{frozen,doolittle,maynard,woese}.
By and large, they have examined the biochemistry and the ease of synthesis
of various biomolecules, and have remained inconclusive due to insufficient
data. I take a different approach in what follows, based on recent advances
in molecular bioinformatics rather than biochemistry. The emphasis is on the
role of biomolecules as building blocks of the genetic languages, i.e. how
efficiently the biomolecules can implement the task required of the languages.
This emphasis on the purpose of the process, and not merely biochemical
synthesis, provides evidence that the present genetic languages arose by
duplication from a simpler form containing 10 amino acids and a doublet
genetic code.

\noindent{\bf Aminoacyl-tRNA synthetases (aaRS):}
The language of genes is translated in to the language of proteins by the
adaptor molecules of tRNA. The truly bilingual molecules in this process
are the aaRS, which attach an appropriate amino acid at one end of the
tRNA molecule corresponding to the anticodon present at the other end.
The genetic code is degenerate, and several different tRNA molecules (with
different anticodons) supply the same amino acid. But the aaRS are unique,
only one for each amino acid. It has been discovered that the aaRS belong
to two distinct classes of 10 members each \citep{eriani,schimmel}.
The two classes clearly differ from each other in sequence and structural
motifs, in active sites and in the position where they attach the amino
acid to the tRNA molecule \citep{moras,lewin}.
This has led to the conjecture that the two classes evolved independently,
and early forms of life could have existed with proteins made up of only
10 amino acids of one class or the other.

\noindent{\bf aaRS-tRNA binding:}
A class-I aaRS binds the acceptor helix of tRNA from its minor groove side,
while a class-II aaRS binds the acceptor helix from its major groove side.
The bound aaRS-tRNA complexes for the two classes thus look like mirror
images of each other \citep{moras}.
The bound complexes have open extended conformations in case of class-I,
while they have closed compact conformations in case of class-II.
Several of the class-I aaRS at times fail to identify their cognate
amino acid and attach a wrong amino acid to the tRNA; subsequent editing
mechanisms restore proper acylation by deacylating misacylated tRNAs
\citep{moras}.
These properties hint that class-I amino acids entered the language of
proteins at a later stage. It is also possible to map the signature
motifs of aaRSs from the two classes, within their catalytic domains,
to a head-to-tail sequence complementarity. This has led to the hypothesis
that the two could have been encoded by complementary strands of the same
ancestral gene \citep{rodin}.

\noindent{\bf Amino acid R-group sizes:}
The backbone of polypeptide chains consists of identical repetitive
units, while the side-chain R-groups of amino acids dictate how the chain
twists and folds to yield proteins of various shapes and sizes. Different
amino acids are labeled according to the chemical properties of their
R-groups, e.g. polar vs. non-polar, aliphatic vs. ring/aromatic, positive
vs. negative charge \citep{lehninger}.
Table 1 demonstrates that each R-group property is equally divided amongst
the two amino acid classes. In addition, for every property, the amino acids
with larger R-groups belong to class-I, while those with smaller R-groups
belong to class-II. Thus the class label for amino acids is a clear binary
code for the size of their R-groups \citep{carbon}.
(The molecular weights in Table 1 indicate the size of the R-groups. Note
that Asn is a shorter side chain version of Gln, and His has a positively
charged R-group but it is close to being neutral. The binary code is not
at all evident if one looks at only the sizes of R-groups, without first
separating the amino acids according to the properties of R-groups.)
This binary code has unambiguous structural significance for packing of
proteins. When an aperiodic polypeptide chain folds in to a compact structure,
cavities of different shapes and sizes are left behind. The use of large
R-groups to fill big cavities and small R-groups to fill small ones can
produce a dense compact structure; indeed proteins have a packing fraction
similar to that for the closest packing of identical spheres.

\begin{table}
\caption{Amino acid properties}
\begin{tabular}{|l|l|r|r|l|}
\hline
Amino & R-group  & Mol.   & Class & Secondary  \\
acid  & property & weight &       & propensity \\
\hline
Gly &           & 75  & II & turn     \\
Ala &           & 89  & II & $\alpha$ \\
Pro & Non-polar & 115 & II & turn     \\
Val & aliphatic & 117 & I  & $\beta$  \\
Leu &           & 131 & I  & $\alpha$ \\
Ile &           & 131 & I  & $\beta$  \\
\hline
Ser &           & 105 & II & turn     \\
Thr &           & 119 & II & $\beta$  \\
Asn & Polar     & 132 & II & turn     \\
Cys & uncharged & 121 & I  & $\beta$  \\
Met &           & 149 & I  & $\alpha$ \\
Gln &           & 146 & I  & $\alpha$ \\
\hline
Asp & Negative  & 133 & II & turn     \\
Glu & charge    & 147 & I  & $\alpha$ \\
\hline
Lys & Positive  & 146 & II & $\alpha$ \\
Arg & charge    & 174 & I  & $\alpha$ \\
\hline
His &           & 155 & II & $\alpha$ \\
Phe & Ring/     & 165 & II & $\beta$  \\
Tyr & aromatic  & 181 & I  & $\beta$  \\
Trp &           & 204 & I  & $\beta$  \\
\hline
\end{tabular}\\
{Properties of amino acids depend on their side chain R-groups.
Larger molecular weights indicate bigger side chains. The 20 naturally
occurring amino acids are divided in to two classes of 10 each, depending
on the properties of aminoacyl-tRNA synthetases that bind the amino acids
to tRNA \citep{moras,lewin}.
The dominant propensities of amino acids for forming secondary protein
structures are also listed \citep{creighton}.}
\end{table}

\noindent{\bf Secondary structure propensities:}
From an evolutionary point of view, the smaller and simpler R-groups of
class-II amino acids are likely to have emerged earlier than the more
elaborate ones of class-I. With the knowledge of a large number of protein
structures, the propensities of individual amino acids to participate in
various secondary structures (i.e. $\alpha$-helices, $\beta$-sheets and
turns) have been identified \citep{creighton}.
In particular, turns are crucial for polypeptide chains to fold in to
compact shapes and produce a variety of structures. Table 1 shows that all
the amino acids with high preferences for turns (Gly, Pro, Ser, Asn, Asp)
belong to class-II. Also, together with other members of class-II, they
are capable of forming all the secondary protein structures.

\noindent{\bf Optimal 3-dim structural language:}
Tetrahedral geometry provides the simplest discrete language that can
encode arbitrary 3-dim structures (in the same manner as Boolean logic
provides the simplest discrete language for 1-dim sequences of letters),
and carbon is the unique element at the atomic scale for its physical
realization \citep{carbon}.
To have the maximum versatility while a polypeptide backbone folds on
a diamond lattice, this language requires 9 orientation instructions per
amino acid building block. These 9 orientations have a good overlap with
the allowed regions of the Ramachandran map (i.e. 3 values each for angles
$\phi$ and $\psi$). Each amino acid class has a special member involved
in transformations beyond these 9 orientations (Cys of class-I forms
long distance disulfide bonds, and Pro of class-II helps in ``trans-cis''
switch).

\noindent{\bf Patterns in the genetic code:}
The triplet genetic code is degenerate. In particular, the third base of
the codon carries only a limited (either binary or none) meaning instead
of four-fold possibilities. This feature, labeled wobble rules \citep{wobble},
is exact for the mitochondrial genetic code. Table 2 shows that the
pyrimidines U,C are equivalent in the third position of the codon, and so
are the purines A,G. This redundancy reduces the codon possibilities to 32,
NNY (Y=U or C) and NNR (R=A or G). A closer inspection of Table 2 shows
that all class-II amino acids, except Lys, can be coded by the codons NNY.
A doublet genetic code NNY, with the third base representing only a
punctuation mark as shown in Table 3, would therefore suffice to encode
the class-II amino acids.

\begin{table}
\caption{Mitochondrial genetic code}
\begin{tabular}{|l|l|l|l|}
\hline
{\bf UUU Phe} & {\bf UCU Ser} & UAU Tyr & UGU Cys \\
{\bf UUC Phe} & {\bf UCC Ser} & UAC Tyr & UGC Cys \\
UUA Leu & {\bf UCA Ser} & UAA Stop& UGA Trp \\
UUG Leu & {\bf UCG Ser} & UAG Stop& UGG Trp \\
\hline
CUU Leu & {\bf CCU Pro} & {\bf CAU His} & CGU Arg \\
CUC Leu & {\bf CCC Pro} & {\bf CAC His} & CGC Arg \\
CUA Leu & {\bf CCA Pro} & CAA Gln & CGA Arg \\
CUG Leu & {\bf CCG Pro} & CAG Gln & CGG Arg \\
\hline
AUU Ile & {\bf ACU Thr} & {\bf AAU Asn} & {\bf AGU Ser} \\
AUC Ile & {\bf ACC Thr} & {\bf AAC Asn} & {\bf AGC Ser} \\
AUA Met & {\bf ACA Thr} & {\bf AAA Lys} & AGA Stop\\
AUG Met & {\bf ACG Thr} & {\bf AAG Lys} & AGG Stop\\
\hline
GUU Val & {\bf GCU Ala} & {\bf GAU Asp} & {\bf GGU Gly} \\
GUC Val & {\bf GCC Ala} & {\bf GAC Asp} & {\bf GGC Gly} \\
GUA Val & {\bf GCA Ala} & GAA Glu & {\bf GGA Gly} \\
GUG Val & {\bf GCG Ala} & GAG Glu & {\bf GGG Gly} \\
\hline
\end{tabular}\\
{The (vertebrate) mitochondrial genetic code differs slightly from the
universal genetic code. The wobble rules are exact for the mitochondrial
code, so the third codon position has only a binary meaning. Class II
amino acids are indicated by boldface letters.}
\end{table}

\noindent{\bf Operational RNA code of the tRNA acceptor stem:}
In the tRNA molecule, the anticodon and the amino acid attachment site
are separated by a distance of $\approx 75$\AA---too far apart for any
direct interaction. It has been observed that the acceptor stem sequence,
which closely interacts with the amino acid, plays a key role in proper
aminoacylation of tRNA. This operational RNA code is formed by the
first four base pairs and the unpaired base N$^{73}$ of the acceptor
stem \citep{secondcode}.
Just the sequence of these bases does not fully describe the operational code,
and there is a substantial variation in base sequences amongst different
living organisms. The operational code actually relies on explicit
structure-dependent atomic recognition between nucleotide bases and amino
acids, where chemical groups and conformational changes play a crucial role.
Still, by examining a large number of tRNA sequences from a wide variety
of living organisms, a common consensus acceptor stem has been constructed.
The consensus sequence shows patterns in the first three base pairs,
that correlate with the anticodon-codon base pairs \citep{rodin2}.
Specifically, (a) the first base pair is almost invariably G$^1$-C$^{72}$
and is mapped to the wobble position, while (b) the second base pair is
mostly G$^2$-C$^{71}$ or C$^2$-G$^{71}$ which correlate well respectively
with the pyrimidines Y and purines R in the middle position of the codons.
Based on these patterns, it has been proposed that the modern tRNAs arose
from repetitive extensions and complementary pairing of short palindromic
acceptor stem sequences \citep{rodin2},
where (i) the 1-2-3 position bases became the forerunners of the anticodons,
and (ii) the G-C rich sequence expanded from G to R and from C to Y.

\begin{table}
\caption{Probable doublet genetic code}
\begin{tabular}{|l|l|l|l|}
\hline
UUY Phe & UCY Ser & UAY --- & UGY --- \\
CUY --- & CCY Pro & CAY His & CGY Lys/Arg \\
AUY --- & ACY Thr & AAY Asn & AGY Ser \\
GUY --- & GCY Ala & GAY Asp & GGY Gly \\
\hline
\end{tabular}\\
{The probable NNY doublet genetic code that evolved to the present
triplet genetic code. Allowing G-U wobble pairing, the anticodons
could be just $\overleftarrow{\rm GNN}$. Note that the code includes
the four possibilities where the two codon bases are identical.}
\end{table}

\noindent{\bf Optimal database search algorithm:}
Syntheses of DNA, RNA and proteins are assembly operations that construct
desired biomolecules by arranging their fundamental building blocks in a
precise manner specified by a template. Prior to assembly, the building
blocks are randomly floating around in the cellular environment, so the
assembly process involves unsorted database search. The search takes place
through molecular bond formation of the building blocks with the pre-existing
template, which is a binary oracle (either it happens or it does not). Also,
natural selection is expected to guide the assembly process towards its
optimal realization. Explorations in computer science has discovered the
optimal assembly algorithm for a binary oracle. It is based on dynamics of
waves and predicts a specific relation between the number of search queries
and the number of items in the database \citep{grover}.
Its predictions match the number of building blocks involved in genetic
languages (i.e. $N=4,10,20$ for $Q=1,2,3$ respectively \citep{quant_gc}),
when a query is identified with nucleotide base-pairing; no other
purposeful explanation of these numbers is known. Explicitly, DNA/RNA
have an alphabet of 4 nucleotide bases identified with 1 base pairing,
polypeptides have an alphabet of 20 amino acids identified by 3 base
pairings, and a single class of 10 amino acids can be identified by
2 base pairings.

\noindent{\bf Exceptions to the global features:}
Taken individually, there are minor variations in the features listed above.
But even they display illuminating patterns:\\
(a) Pro and Cys, belonging to class-II and class-I respectively, are oddballs
in the tetrahedral structural language. In some archaea (e.g. M. jannaschii
and M.  thermoautotrophicum), ProRS synthesizes both Pro-tRNA$^{\rm Pro}$ and
Cys-tRNA$^{\rm Cys}$. This is the only known example of dual functionality
amongst all the aaRS \citep{woese},
perhaps a relic of the doubling phenomenon.\\
(b) The only class-II amino acid not coded by NNY codons is Lys, while
its class-I counterpart Arg can be coded by the NNY codons. Lys is the
only amino acid having two distinct aaRS, one belonging to class-I (in
most archaea) and the other belonging to class-II (in most bacteria and
all eukaryotes) \citep{woese}.
On the other hand, ArgRS is the most complex of all the aaRS,
with a large diversity amongst organisms \citep{woese}.
This may be an indication of an exchange of roles having occurred between
Lys and Arg in the genetic machinery.\\
(c) All the variations seen in the genetic code (i.e. differences in
nuclear and mitochondrial codes, and locations of seleno-cysteine and
Stop codons) are of the NNR type \citep{lewin},
consistent with the inference that NNR codons were roped in the genetic
code later.\\
(d) The operational RNA code is G-C rich, but it also has frequent
occurrences of G-U wobble pairs. Allowing for this wobble pairing,
the anticodons of NNR codons could just be $\overleftarrow{\rm GNN}$.
The anticodons $\overleftarrow{\rm GNN}$, with N restricted to G or C or U
(but not A), cover 9 of the 10 class-II amino acids in Table 3.
The exception is Phe with anticodon $\overleftarrow{\rm GAA}$, and
PheRS-tRNA binding is the only case where a class-II aaRS attaches amino
acid to ribose with the stereochemistry of a class-I aaRS \citep{moras}.

\bigskip
The features described above span three different types of molecules
involved in genetic information processing---aminoacyl-tRNA synthetases,
amino acids and nucleotide bases. Individually the features may be brushed
aside as chance occurrences, but as a whole they form a tightly woven web
that cannot be ignored. Indeed, there is a strong case for the following
evolution of the genetic information processing machinery:
The ancestors to the present proteins were those synthesized from only the
10 class-II amino acids. These 10 amino acids were encoded by the operational
RNA code on the acceptor stem of the tRNA, read from the major groove side.
This operational genetic code can be effectively interpreted as the doublet
form NNY shown in Table 3, with anticodons $\overleftarrow{\rm GNN}$.
Quite likely, the migration of the doublet code from the acceptor stem to
the anticodon site of tRNA allowed the set of 10 amino acids to be doubled
by including larger R-groups for every property. The opportunity for
expansion of the code was provided by its transition from the paired
bases on the acceptor stem to the unpaired anticodon. On the other hand,
the motivation for expansion came from the improved packing of proteins,
without disrupting previously established structures, by filling up some
of the unfilled large cavities that existed. This expansion took different
routes for the operational RNA code of the tRNA acceptor stem and the
genetic code of the anticodons, leading to their divergence. The doubled
set of amino acids were included in the acceptor stem code by allowing
aaRS-tRNA binding from the minor groove side, while they were included in
the anticodons by converting its third base from a punctuation mark to a
binary value. More frequent inclusion of A-U in the G-C rich anticodons,
and subsequent refinements, brought the genetic code to its present universal
form.

The above described ``doubling of the genetic code'' hypothesis firmly
shifts the evolutionary emphasis from ``frozen accident'' \citep{frozen}
to ``optimal solution''. This is not the first time that a primitive doublet
genetic code has been proposed. But now its form is explicit, and the inputs
leading to it have arisen from a different perspective. The sole purpose of
the genetic machinery, apart from self-sustenance, is reliable and efficient
transmission of hereditary information regarding protein structures.
Considerations of optimal encoding of languages provide a direction to
evolution, that can make specific choices amongst the many possibilities
thrown up by blind biochemistry. As a matter of fact, information theory
and biochemistry (i.e. software and hardware) can complement each other
in narrowing down the multitude of options, while trying to understand
the origin of life.

The evolutionary scenario presented here suggests which class-I amino acid
substituted which class-II amino acid of similar property during doubling.
Some pairs are easy to guess, e.g. (Asn,Gln), (Asp,Glu), (Lys,Arg), (Pro,Cys),
while figuring out others would require a more detailed understanding of
the atomic recognition mechanism in the operational RNA code.
The evolutionary scenario can also be strengthened by finding more
supporting evidence. For example, highly conserved regions of ancestral
proteins should be dominated by class-II amino acids (the two classes of
amino acids are thoroughly mixed in present proteins). Moreover, it should
be possible to make artificial proteins of diverse functionality from the
class-II amino acids alone.

% How well does this match with the known amino acid substitution rules
% for proteins?

All this can only be the first step towards unraveling the mystique
surrounding the origin of life. There are further questions one may ask:\\
(i) ``Why did the translation machinery for the doublet code move in steps
of three bases, even when only two bases carried information, leaving the
third base as a punctuation mark?'' The answer is likely to be found in
the stereochemistry between amino acids and tRNA acceptor stem nucleotide
bases.\\
(ii) ``How did the genetic
machinery involving 10 class-II amino acids come about?'' The optimality
criteria point towards a still earlier predecessor involving 4 amino acids,
perhaps encoded by one pair of complementary nucleotide bases.\\
(iii) ``Is there a relation between the doublet code and the RNA world?''
The doublet code is intimately tied to the properties of amino acids,
and hence has to appear after replacement of ribozymes by polypeptides.
May be it would be easier to construct a mapping between properties of
ribozymes and the smaller set of class-II amino acids.\\
Although conjectures can be made, clear answers to such questions would
require more clues and careful modeling.

% A curious feature is that the doublet code would be flanked by purines
% in the anticodon loop of tRNA, in the palindromic pattern [RNN]R,
% with R likely to be G/inosine.


\begin{thebibliography}{}

\bibitem[Arnez and Moras(1997)]{moras}
         Arnez, J.G., Moras, D., 1997.
{Structural and functional considerations of the aminoacylation reaction}.
         {Trends Biochem.\ Sci.\/} {22}, 211-216.
\bibitem[Crick(1966)]{wobble}
         Crick, F.H.C., 1966.
{Codon-anticodon pairing: The wobble hypothesis}.
         {J.\ Mol.\ Biol.\/} {19}, 548-555.
\bibitem[Crick(1968)]{frozen}
         Crick, F.H.C., 1968.
{The origin of the genetic code}.
         {J.\ Mol.\ Biol.\/} {38}, 367-379.
\bibitem[Eriani et al.(1990)]{eriani}
         Eriani, G., Delarue, M., Poch, O., Gangloff, J., Moras, D., 1990.
{Partition of tRNA synthetases in to two classes based on mutually
         exclusive sets of sequence motifs}.
         {Nature\/} {347}, 203-206.
\bibitem[Grover(1996)]{grover}
         Grover, L.K., 1996.
{A fast quantum mechanical algorithm for database search}.
         In: {Proceedings of the 28th annual ACM symposium
         on theory of computing\/}, Philadelphia, pp.212-219.
         {\tt arXiv.org:quant-ph/9605043}.
\bibitem[Lehninger et al.(1993)]{lehninger}
         Lehninger, A.L., Nelson, D.L., Cox, M.M., 1993.
         {Principles of biochemistry},
         second edition. Worth publishers, USA.
\bibitem[Lewin(2000)]{lewin}
         Lewin B., 2000. {Genes VII}.
         Oxford Univ. Press, Oxford.
\bibitem[Maynard Smith and Szathm\'ary(1995)]{maynard}
         Maynard Smith, J., Szathm\'ary, E., 1995.
         {The major transitions in evolution}.
         W.H. Freeman, Oxford.
%        {\it The origins of life: From the birth of life to the origin
%        of language\/} (Oxford Univ. Press, Oxford, 1999).
\bibitem[Patel(2001)]{quant_gc}
         Patel, A., 2001.
{Why genetic information processing could have a quantum basis}.
         {J.\ Biosc.\/} {26}, 145-151.
         {\tt arXiv.org:quant-ph/0105001}.
\bibitem[Patel(2002)]{carbon}
         Patel, A., 2002.
{Carbon---the first frontier of information processing}.
         {J.\ Biosc.\/} {27}, 207-218.
         {\tt arXiv.org:quant-ph/0103017}.
\bibitem[Rodin and Ohno(1995)]{rodin}
         Rodin, S.N., Ohno, S., 1995.
{Two types of aminoacyl-tRNA synthetases could be originally encoded by
complementary strands of the same nucleic acid}.
         {Origins Life Evol.\ Biosphere\/} {25}, 565-589.
\bibitem[Rodin, Rodin and Ohno(1996)]{rodin2}
         Rodin, S, Rodin, A., Ohno, S., 1996.
{The presence of codon-anticodon pairs in the acceptor stem of tRNAs}.
         {Proc.\ Nat.\ Acad.\ Sci.\ USA\/} {93}, 4537-4542.
\bibitem[Schimmel(1991)]{schimmel}
         Schimmel, P., 1991.
{Classes of aminoacyl-tRNA synthetases
and the establishment of the genetic code}.
         {Trends Biochem.\ Sci.\/} {16}, 1-3.
\bibitem[Schimmel et al.(1993)]{secondcode}
         Schimmel, P., Giege, R., Moras, D., Yokoyama S., 1993.
{An operational RNA code for amino acids
and possible relationship to genetic code}.
         {Proc.\ Nat.\ Acad.\ Sci.\ USA\/} {90} 8763-8768.
\bibitem[S\"oll and Doolittle(1995)]{doolittle}
         S\"oll, D., Doolittle, R. (Eds.), 1995.
         {Special issue: The aminoacyl-tRNA synthetases
         and the evolution of the genetic code}.
         {J.\ Mol.\ Evol.\/} {40(5)}.
\bibitem[Creighton(1993)]{creighton}
         Creighton T.E., 1993. {Proteins: Structures and molecular properties},
         second edition. W.H. Freeman, New York.
\bibitem[Woese et al.(2000)]{woese}
         Woese, C.R., Olsen, G.J., Ibba, M., S\"oll, D., 2000.
{Aminoacyl-tRNA synthetases, the genetic code, and the evolutionary process}.
         {Microbio. and Mol. Bio. Rev.\/} {64}, 202-236.

\end{thebibliography}
\end{document}